\documentclass[lp,twocolumn]{jpsj3}
\usepackage{txfonts}

\usepackage{amssymb,amsmath}
\allowdisplaybreaks
\usepackage{mathtools}
\usepackage{bm}
\usepackage{physics}

\newcommand{\sgn}{\mathrm{sgn}}        
\newcommand{\vNF}{{\bm{N}}}
\newcommand{\vNG}{{\bm{n}}}
\newcommand{\kF}{{\kappa}}
\newcommand{\kG}{{\kappa '}}
\newcommand{\kP}{{K}}
\newcommand{\kD}{{K}}
\newcommand{\aF}{{a}}
\newcommand{\aG}{{a '}}
\newcommand{\aP}{{A}}
\newcommand{\aD}{{A}}
\newcommand{\bF}{{b}}
\newcommand{\bG}{{b '}}
\newcommand{\eq}{{\mathrm{eq}}}

\newcommand{\phiF}{{\phi}}
\newcommand{\phiG}{{\psi}}
\newcommand{\dphiF}{{\dot{\phi}}}
\newcommand{\dphiG}{{\dot{\psi}}}
\newcommand{\ddphiF}{{\ddot{\phi}}}
\newcommand{\ddphiG}{{\ddot{\psi}}}

\newcommand{\LF}{{\mathcal{L}_q}} 
\newcommand{\LP}{{\mathcal{L}_p}} 
\newcommand{\GF}{{G_q}} 
\newcommand{\GP}{{G_p}} 

\newcommand{\vr}{{\bm{r}}}

\title{Dynamics of Composite Domain Walls in Multiferroics 
in Magnetic Field \\
and Their Instability}

\author{
Koji Kawahara 
and 
Hirokazu Tsunetsugu
}

\inst{%
The Institute for Solid State Physics, 
The University of Tokyo, 
Kashiwanoha 5-1-5, Chiba 277-8581, Japan
}
\date{August 19, 2020}

\abst{
We study theoretically the dynamics of 
composite domain walls (DW) in multiferroic
material GdFeO$_3$  driven by magnetic field $H$.
Two antiferromagnetic orders of Fe and Gd spins interact 
Gd-ion displacement in this system with coupling $c$ at low temperatures, 
and we have numerically simulated the corresponding time-dependent
Ginzburg-Landau equations in which magnetic field $H$ 
couples to Fe-spin order parameter.  
We vary $H$ and $c$ systematically and calculate velocity $v$ 
and inner structure in a stationary state for
magnetic DW and magneto-electric DW.
DW mobility $v/H$ increases with $H$ and decreases with $c$ for both DWs,
but their characteristics differ between the two.  
We have also studied analytically the smooth characteristics
of magneto-electric DW by a perturbation theory.  
Another finding is a splitting instability at large $H$.
A magneto-electric DW splits into a pair of magnetic DW and electric DW
when $c$ is large, while a magnetic DW splits when $c$ is small.
Internal structure of composite DW deforms with increasing $c$ and $H$,  
and modulations in different order parameters separate in space.
Their relative distances show a noticeable 
enhancement with approaching the splitting instability. 
}

\begin{document}
\maketitle


\section{Introduction}
\label{sec1:intro}
Multiferroic materials are a system in which multiple macroscopic
orders coexist, and a rich play ground of physics of both fundamental
and application interests \cite{mf1,mf2,mf3,mf4}.
They exhibit complex low-temperature ordered phases. 
Extensive researches have been performed 
for clarifying microscopic origin of multiferroic phases \cite{Tokura2014} 
and also symmetry conditions \cite{Perez-Mato2016}.  
The multiferroic phases have domain walls (DWs) 
as a topological defect \cite{defect}, 
and their internal structure is 
affected by interactions between multiple order parameters.
These DWs are often called {\it composite domain wall} \cite{Tokunaga2009},
when several order parameters ``flip'' around the same place. 
Low-temperature phases and their DW excitations 
have been experimentally studied in detail for several
multiferroics such as 
TbMnO$_3$ \cite{Kimura2003}, 
GdFeO$_3$ \cite{Tokunaga2009}, 
and $R$MnO$_3$ \cite{Choi2010,Ishiwata2010}. 
As for application interests,  multiferroics is a candidate of new
low-energy consuming memory device using cross 
correlations \cite{bibes,Belashchenko2016}.
For example, it is possible to flip or reorient magnetic moments 
by applying electric field. 
This effect originates in either reorientation 
of bulk spins \cite{Kagawa2009}, 
or creeping motion of composite DW \cite{Tokunaga2012}. 

An important issue is how the interaction between different orders
affects the dynamics of the system. Multiple orders 
in multiferroic materials have their intrinsic time
and length scales that differ from each other.  
When their interaction takes effect,  
the dynamics of the coupled system becomes non-trivial,
and it is important to understand their characteristics.  
The DW dynamics in a system with single order parameter 
is quite well understood.
Bloch DW in uniaxial magnet is driven by magnetic field 
and its dynamics is described by Walker's 
solution \cite{Schryer1974}, 
which predicts a linear velocity-field characteristic. 
A recent study also discussed a turbulent DW motion 
at high field \cite{beach}. 
It is also known that 
electric current drives magnetic DW 
by spin transfer effect~\cite{Berger1973,Tatara}.  
For multiferroics, there are several studies 
on their bulk dynamics as well as static DW.
Optical responses were studied based on a spin model \cite{Kagawa2009} 
and also first principal calculation \cite{Zhu2017}. 
Spin reorientation process of bulk spin 
under magnetic field was experimentally visualized 
and analyzed by micromagnetic simulation of 
Landau-Lifshitz-Gilbert equation \cite{Matsubara2015}.
Spatial structure of static composite DWs was 
studied using a Ginzburg-Landau model 
for magneto-electric multiferroic materials such as 
BiFeO$_3$ \cite{Scott2010} 
and also ferroelectric-ferroelastic DW 
in BiFeO$_3$ \cite{Eliseev2019}. 
The former case has two order parameters. 
The latter case has three order parameters, 
and instability of static DW was examined.  
Several interesting phenomena are found 
such as enhancement of order parameter 
around DW \cite{Scott2010} 
and transverse deformation of DW \cite{Eliseev2019}.

In this paper, we will investigate dynamics of composite DW,
and the main target material is GdFeO$_3$.
In this system, three order parameters interact at low temperatures, 
and they are two antiferromagnetic orders and lattice distortion. 
One antiferromagnetic order is made of Fe spins and 
staggered Dzyaloshinsky-Moriya interactions \cite{DM1,DM2}
generate parasitic weak ferromagnetic moments.
Ferroelectricity originates in displacement of Gd
ions induced by interactions with antiferromagnetic orders 
of Fe and Gd spins, and this appears at $T < 4K$.  
Thus, this is one of the typical multiferroic materials
with both ferromagnetism and ferroelectricity \cite{Tokunaga2009,Mukhin2004}.
Composite DWs differ in one important point from DWs 
in the systems with single order parameter, 
and that is multiple types of composite DWs depending on
which order parameters change.
Since the low-temperature phase of GdFeO$_3$ has three types
of order parameters, 
there are three types of composite DWs \cite{Tokunaga2009}.  

The main issue to study in this work is how the coupling $c$ 
between the three order parameters changes 
the internal structure and dynamics of composite DWs.  
A fundamental question is how the coupling changes velocity 
of DW driven by magnetic field $H$. 
By simulating phenomenological time-evolution equations, 
we study this dynamics and calculate the DW velocity 
in the stationary state with varying field $H$ and coupling $c$.
We will show that increasing $H$ not only
accelerates the velocity of composite DWs but also destabilizes them.  
It is interesting that this instability depends on their DW type
as will be shown later.
We also perform an analytical calculation for correction 
in DW velocity due to the coupling, when $c$ and $H$ are both small.  
This DW instability is also related to another question
about the inner structure of composite DWs.
We will calculate spatial profile of the multiple order parameters
inside a DW, and find that it also demonstrates a precursor 
of the DW instability.  
For simplicity, we limit our study in this work
to the case that order parameters vary only along one direction of space.
Therefore, transverse thermal fluctuations inside DW region
are neglected, which may be justified at low temperatures.  

This paper is organized as follows. 
In Sec.~\ref{sec2:model}, we explain our phenomenological model 
including the interaction $c$ among multiple order parameters,  
basics of composite DWs, and time-evolution equations to solve. 
In Sec.~\ref{sec3:Numerical}, we numerically solve these equations 
for DWs driven by magnetic field $H$ and calculate their velocity $v$.  
The scaling of $v(c,H)$ is analyzed in detail.  
In Sec.~\ref{sec4:instability}, we investigate an instability 
of DWs based on numerical results, and 
analyze internal structure of DWs.  
In Sec.~\ref{sec5:analytical}, we formulate a perturbative 
theory when the interaction $c$ is small. 
The last section is conclusions.  


\section{Model}
\label{sec2:model}
\begin{figure}[tb]
\centering
\includegraphics[width=1.0\linewidth]{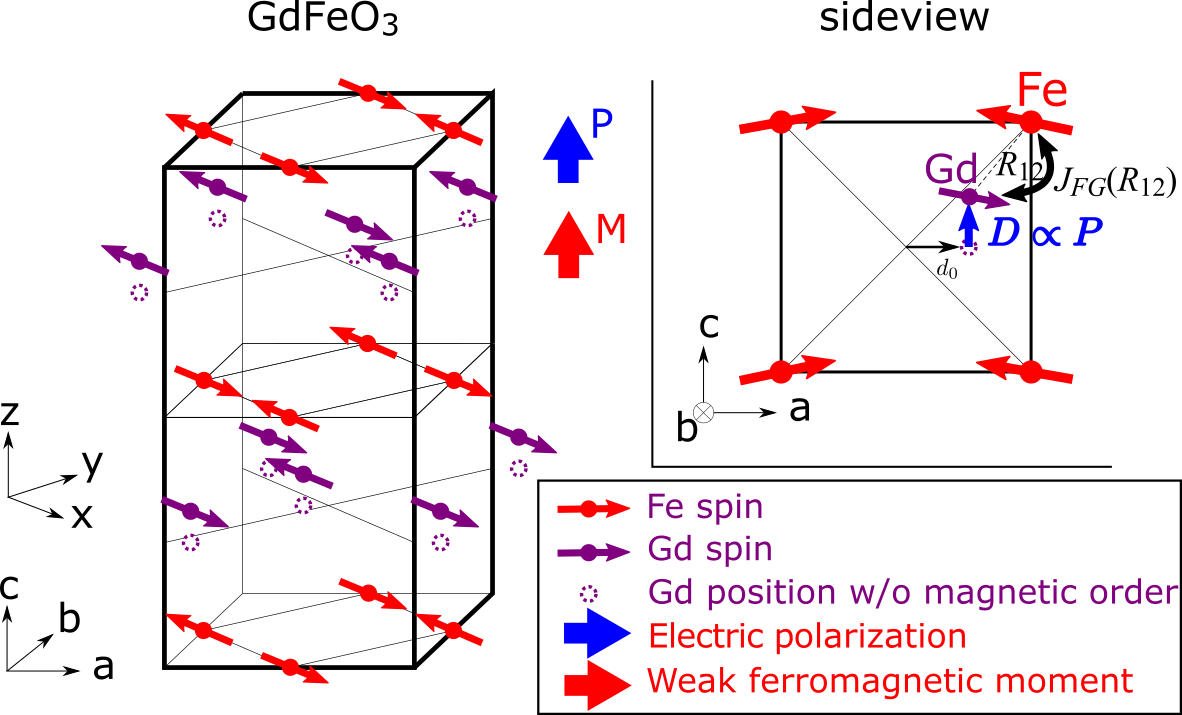}
\caption{
Magnetic unit cell of GdFeO$_3$ with no external field below 4 [K]. 
$a$, $b$, and $c$ denote crystallographic axes, 
while $x$, $y$, and $z$ are quasi-cubic axes. 
}
\label{fig:gdfeo3crystal}
\end{figure}

In this paper, we study the dynamics of composite DW
in the typical multiferroic material GdFeO$_3$.
This has a crystal structure that contains two magnetic 
sublattices comprised of Fe and Gd ions each ,
and exhibits both complex magnetic order and ferroelectricity 
at low temperatures below 4 [K].  
At temperatures $T < 650$ [K], Fe spins show an antiferromagnetic 
order accompanied by weak ferromagnetic order. 
Below the transition at 4 [K], Gd spins show an
antiferromagnetic order, and ferroelectricity appear at the same 
time \cite{Tokunaga2009, Das2017}.
The crystal and magnetic structures are shown
in Fig.~\ref{fig:gdfeo3crystal}.  
Magnetic structure of Fe and Gd spins is the type  
$G_x A_y F_z$ and $G_x A_y$, respectively, 
in Bertaut notation \cite{Bertaut1971}, 
and thus the $z$-component of Fe spins is
weak ferromagnetic moment.
Ferroelectricity is induced by antiferromagnetic interactions 
between Fe and Gd spins via "exchange-striction mechanism", 
which leads to uniform displacement of 
Gd ions \cite{Tokunaga2009}.  
Interactions between each spins are also estimated 
both experimentally \cite{Cashion1970} 
and by first principle calculation \cite{Zhu2017}.
 

\subsection{Ginzburg-Landau free energy}
\label{sec21:GL}
We employ a coarse-grained model instead of microscopic spin
Hamiltonian to study the dynamics of composite DW, 
since it is more convenient to study a large-size object.  
That is the Ginzburg-Landau free energy model for GdFeO$_3$ 
and it is represented in terms of three continuous field variables: 
local antiferromagnetic moments of Fe and Gd spins 
$\vNF (\vr )$ and $\vNG (\vr )$ 
and ferroelectric polarization density $P (\vr)$
\begin{subequations}
\begin{align}
&{F} = F \bigl[ \vNF (\vr ) , \vNG (\vr ) , P (\vr ) \bigr] = 
  \int d\vr \, \mathcal{F} (\vr ),
\label{eq:total_F1}  
\\
&\mathcal{F} = \mathcal{F}_0 + \mathcal{F}_1 , 
\ \
\mathcal{F}_0
= \mathcal{F}_F +  \mathcal{F}_G + \mathcal{F}_P ,
\label{eqn:free_ene_3D}
\end{align}
\label{eqs:total_F}
\end{subequations}
where each part of the order parameters reads
\begin{subequations}
\begin{align}
&\mathcal{F}_F =
\frac{\kF }{2} \qty(\nabla \vNF)^2 
-\frac{\aF}{2}\abs{\vNF}^2 +\frac{\bF}{4}\abs{\vNF}^4 
+\sum_{\nu=x,y,z} \frac{\alpha_\nu}{2}(N_\nu)^2 - H N_x  , 
\\
&\mathcal{F}_G =
  \frac{\kG }{2}\qty(\nabla \vNG)^2 
-\frac{\aG}{2}\abs{\vNG}^2 +\frac{\bG}{4}\abs{\vNG}^4  , 
\\
&\mathcal{F}_P =
 \frac{\kD }{2} \qty(\nabla P)^2 + \frac{\aD}{2} P^2 . 
\end{align}
\label{eqs:Fs}
\end{subequations}
The interaction part $\mathcal{F}_1$ will be explained later. 
Both of $\vNF$ and $\vNG$ are 3-component vector fields, and 
we will sometimes use the polar representation 
\begin{align}
\vNF 
&= |\vNF | \, 
(\sin \theta_F \cos \phi , 
 \sin \theta_F \sin \phi , 
 \cos \theta_F) , 
\nonumber\\
\vNG 
&= |\vNG | \, 
(\sin \theta_G \cos \psi , 
 \sin \theta_G \sin \psi , 
 \cos \theta_G) . 
\label{eqn:polarrep}
\end{align}
Electric polarization $P$ originates in the $z$-component 
of Gd ion displacement.  
$\alpha_\nu$'s are onsite biaxial anisotropy constants 
of Fe spins ($x$-easy, $z$-hard); 
$\alpha_x <  \alpha_y  <  \alpha_z =0$, 
while Gd spins are isotropic \cite{Tokunaga2012}. 

Magnetic field $H$ couples to the $x$-component 
of staggered moment of Fe spins. 
Of course, this is not a direct Zeeman coupling but an effective one. 
The staggered component of Fe-spin moments couples to the uniform 
component through staggered Dzyaloshinsky-Moriya 
interactions \cite{Tokunaga2009,DM1,DM2}. 
Since uniform magnetic field $\bm{h}$ induces the uniform component, 
the coupling has a form 
$- (\bm{D}^{\mathrm{stag}} \times \chi \bm{h}) \cdot \bm{N}$ 
where $\chi$ is the magnetic susceptibility tensor. 
When $\bm{h}$ is applied along $z$-axis and $y$-component is active 
in $\bm{D}^{\mathrm{stag}}$, this form is reduced to $-H N_x$ with 
$H=D^{\mathrm{stag}}_y \chi_{zz} h_z$.  
In our model, we use the induced effective staggered field $H$, 
instead of bare uniform field $h_z$ for simplicity. 

The interaction has a multi-linear form of the three order 
parameters \cite{Tokunaga2009} 
\begin{equation}
\mathcal{F}_1 = c P \vNF \cdot \vNG  .  
\label{eqn:GLfreeenergy}
\end{equation}
One can show this form of coupling based on symmetry argument, 
but we will examine a condition of $c \ne 0$ starting from a microscopic model. 
The starting model is a Heisenberg model 
in which the antiferromagnetic 
coupling $J_{FG} (R_{12})$ between nearest-neighbor Gd and Fe spins 
varies with their distance $R_{12}$. 
This implies that $J_{FG}$ modulates in space with the displacement 
vector $\bm{D} (\vr )$ of Gd ions.  
Assuming the isotropic proportionality
$ \bm{P} (\vr ) = \zeta \bm{D} (\vr )$, 
the result of leading-order coupling is written as 
\begin{equation}
 c = g_{12} \, \zeta^{-3} \, 
\bigl\langle \!\! \bigl\langle P_x (\vb*{R} ) P_y (\vb*{R} ) 
\bigr\rangle \!\!  \bigr\rangle_{\mathrm{av}}, 
\end{equation}
where the average is taken over all the Gd sites $\vb*{R}$. 
The prefactor is given by 
\begin{equation}
  g_{12} = 
R^{-3} \, 
\bar{\nabla} (\bar{\nabla} - 2 ) (\bar{\nabla} - 4) 
J_{FG} (R ) \Bigr|_{R=\bar{R}_{12}},  \ \ \ 
\end{equation}
where $\bar{R}_{12}$ is the average nearest-neighbor distance between
Fe and Gd sites and the dimensionless derivative is defined as
$\bar{\nabla} = d/d \, \log R $.
Therefore, Fe and Gd antiferromagnetic moments couple to each other 
mediated by a ferro quadrupole of Gd-ion displacements.  
This quadrupole has a nonzero amplitude 
when Gd-ion displacements have a staggered order 
in their $x$ and $y$ components. 
This is the case of GdFeO$_3$, but the above condition 
is satisfied by many other situations. 
One example is the case that each of $x$ and $y$ component 
is random in space but their product has a uniform order, 
which is an intrinsic quadrupole order of displacements.  

\subsection{Bulk states and composite domain walls}
\label{sec22:DW}
\begin{figure}[tb]
\centering
\includegraphics[width=0.8\linewidth]{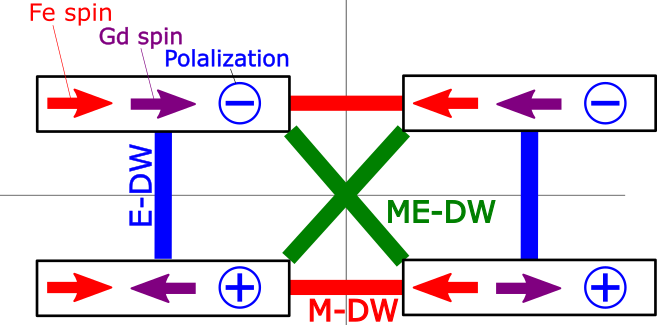}
\caption{
Four degenerate stable states in the model. 
Arrows represent spins along easy axis ($x$-axis), 
and electric polarization is represented by 
the sign of its $z$-component. 
Lines connecting different stable states correspond to 
three types of domain walls; 
electric (E-DW), magnetic (M-DW), and magneto-electronic (EM-DW).}
\label{fig:GrowndStates}
\end{figure}

Stable bulk state is the uniform state with free energy $F$ minimum. 
Since it is a vacuum of topological defects such as DWs, 
it is important to determine it first.  
When no magnetic field is applied $H=0$, 
stable bulk state has 4-fold degeneracy
\begin{subequations}
\begin{align}
  \vNF_{\eq} (\vr ) &= \vNF_{\eq} = \pm (N_{0},0,0),
\label{eq:stable_NF}  
\\  
\vNG_{\eq} (\vr ) &=  \vNG_{\eq}= \pm (n_{0},0,0),  
\label{eq:stable_NG}  
\\
P_{\eq} (\vr ) &= P_{\eq}
=
-\frac{c}{\aD} \vNF_{\eq}\cdot \vNG_{\eq} 
= \pm \frac{c}{\aD} N_0 n_0 , 	
\label{eq:stable_P}  
\end{align}
where 
\begin{align}
\left[ 
\begin{array}{c}
N_{0}^2  \\[4pt] 
n_{0}^2  
\end{array}
\right]
&= 
\frac{1}{ \bF \bG - c^4 / \aD ^2 }
\left[ 
\begin{array}{cc}
 \bG  & c^2 / \aD 
\\[4pt]
c^2 / \aD  & \bF 
\end{array}
\right]
\left[ 
\begin{array}{c}
\aF - \alpha_x   \\[4pt] 
\aG 
\end{array}
\right] . 
\end{align} 
\label{eqs:bulksols}
\end{subequations}
Note that the sign of $P_{\eq}$ is determined from 
$\vNF _{\eq} \cdot \vNG _{\eq}$.
Both $\vNF$ and $\vNG$ point to $x$-axis,  the easy axis of Fe spins.  
These four stable states are shown in Fig.~\ref{fig:GrowndStates}.

DWs have a finite excitation value of free energy
and connect different stable states.
In GdFeO$_3$, two of the three order parameters continuously flip
in DW as pointed out by Tokunaga et al.~\cite{Tokunaga2009}. 
Flipping one order parameter is energetically prohibited, because 
it costs not local but bulk energy of the interaction term $\mathcal{F}_1$.  
Depending on which two are flipped, 
DWs are categorized into three types shown in Fig.~\ref{fig:GrowndStates}.

The first type is magnetic domain wall (M-DW), in which 
$\vNF$ and $\vNG$ flip and correspondingly 
the weak ferromagnetic moment also flips.  
The second type is electric domain wall (E-DW),  
in which $\vNG$ and electric polarization $P$ flip. 
The third type is magneto-electric domain walls (ME-DW), in which 
$\vNF$ and $P$ flip.
This time the weak ferromagnetic moment $M$ and electric polarization $P$ 
flip simultaneously.
In each of these three types, at least one antiferromagnetic 
order parameter rotate inside a DW during its flipping process.  
The spin anisotropy confines this rotation in the $xy$-plane, and 
this rotation defines its chirality $\pm 1$ 
corresponding to clockwise and anticlockwise rotation. 
Therefore, there are 6 types of DW in total.  


\subsection{Time-dependent Ginzburg-Landau equations}
\label{sec23:TDGL}
We will analyze the dynamics of DWs in the following sections, 
and need to choose equations of motion of the order parameters.  
Since none of the order parameters are conserved quantities, 
we use in this work the time-dependent Ginzburg-Landau 
(TDGL) equation \cite{TDGL}. 
An alternative approach \cite{Belashchenko2016} used 
a Lagrangian formulation for collective coordinates 
of DW and studied precession of sublattice magnetizations.  
In this paper, we focus on dissipation-driven effects 
on DW dynamics and use the TDGL formulation.  
Namely, the velocity of order parameters follows the force 
generated by local free energy.  
The general TDGL formulation implies 
$
\partial_t \mathcal{O}(\vr ,t) 
= -\gamma_\mathcal{O} \, 
\delta  F / 
\delta  \mathcal{O} (\vr ,t)
$
for each order parameter field $\mathcal{O}=\vNF$, $\vNG$, or $P$.
The right-hand side is a functional derivative of the total
free energy (\ref{eq:total_F1}) defined by the order parameters
at time $t$.  
Note that we do not include white noises, which 
are sometimes added to the deterministic forces.  
In the present case, the TDGL equations read
\begin{subequations}
\begin{align}
\gamma^{-1} \partial_t N_\nu (\vr , t)
&= 
(\kappa \nabla^2 +a - \alpha_\nu - b \vNF ^2 ) N_\nu
\nonumber\\
&\ \ \ 
+  H \delta_{\nu,x}
-  c P n_{\nu}
, 
\label{eq:TDGL_F}
\\  
{\gamma'}^{-1}\partial_t n_\nu (\vr , t)
&= 
(\kappa' \nabla^2  +a' -  b' \vNG ^2 ) n_\nu 
- c P N_{\nu} , 
\label{eq:TDGL_G}
\\
\Gamma^{-1}\partial_t P (\vr , t)
&= 
(K \nabla^2 - A ) P 
- c \vNF \cdot \vNG , 
\label{eq:TDGL_P}
\end{align}
\label{eqs:TDGL}
\end{subequations}
for $\nu = x$, $y$, and $z$, 
and $\delta_{\nu ,x}$ is Kronecker delta. 
The coefficients $\gamma$, $\gamma '$,
and $\Gamma$ are phenomenological relaxation constants.  
In the next sections, we will perform both numerical and analytical 
analyses of this set of TDGL equations.  

\begin{figure}[tb]
\centering
\includegraphics[width=\linewidth]{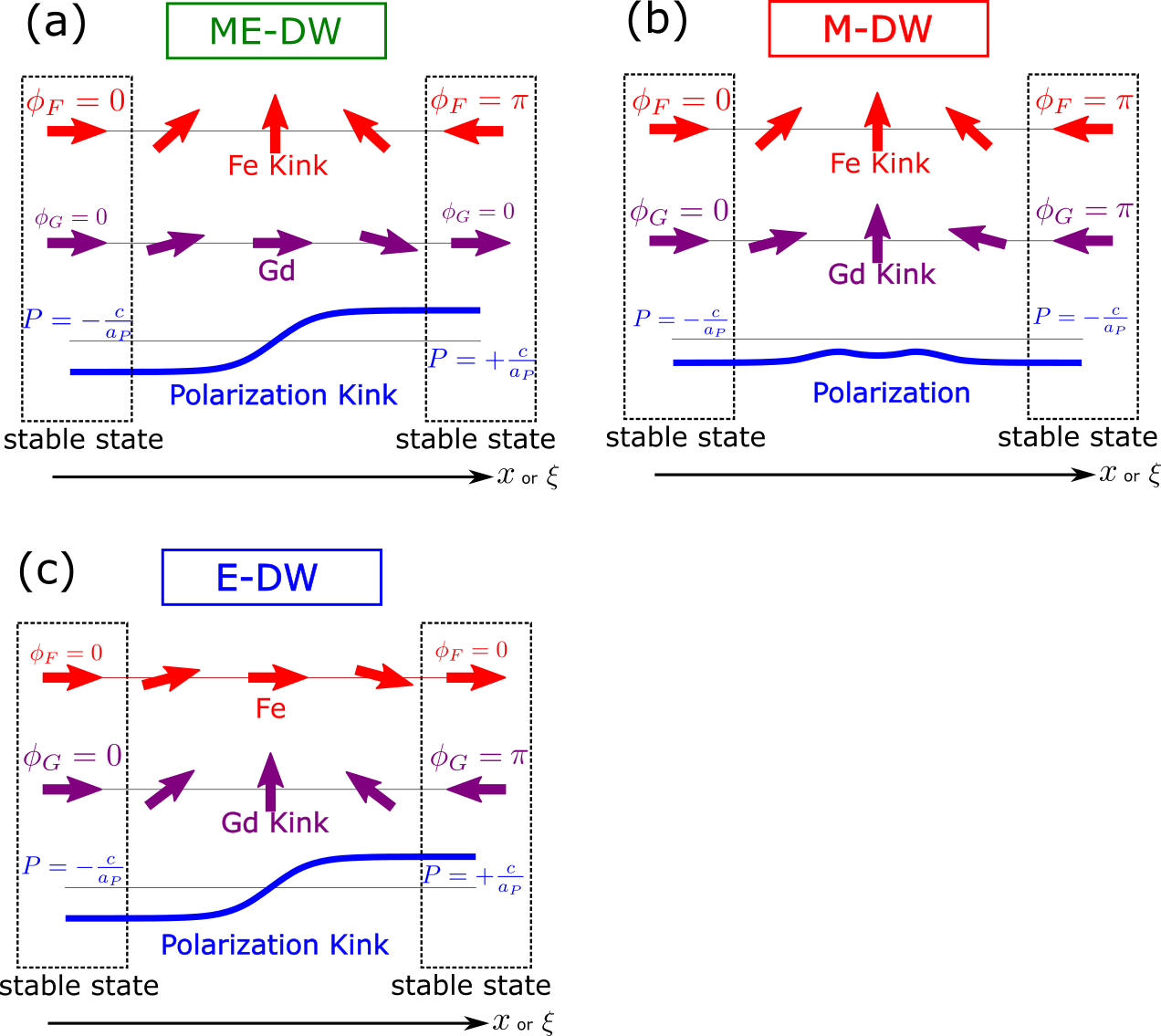}
\caption{
Schematics of inner structures of 
(a) multiferroic (b) magnetic (c)electric DWs driven 
by magnetic $H$ or electric $E$ field. 
Arrows show spin directions 
versus the boosted coordinate $\xi$ moving with domain wall. 
In each DW, two out of the three order parameters flip, 
while the non-flipping order parameter is modulated 
around the DW center.}
\label{fig:medw-schematics}
\end{figure}

Before analyzing the dynamics of composite DWs, let us recall
the result of a simple case of no interaction $c=0$ and
flat DW, \textit{i.e.}, one-dimensional $\vr$-dependence.  
In this case, external field $H$ drives only Fe order parameter $\vNF$. 
When the amplitude modulation is small, 
its stationary state solution is
\begin{subequations}
\begin{align}
  &\vNF (\vr , t) = N_0
  \bigl( \cos \phi_s (\xi ) , \, \sin \phi_s (\xi ) , \, 0 \bigr) ,
  \\
  &\xi = x - \mu_0 H t + x_0 ,
  \ \
  \mu_0 \equiv \gamma N_0 \sqrt{\frac{\kappa}{\alpha_y - \alpha_x }} .
\label{eq:mu0}  
\end{align}
\end{subequations}
The functional form of $\phi_s$ will be given
later in Eq.~(\ref{eq:Walker_sol})
and $x_0$ is a constant determined by the initial condition.
Therefore, the terminal velocity of DW motion
is linear in the applied field $H$
\begin{equation}
  v = \mu_0 H . 
\end{equation}
This means that the coefficient $\mu_0$ is a DW mobility at $c=0$ and
it is proportional to the DW width
$\ell_0 = N_0 \sqrt{\kappa / (\alpha_y - \alpha_x )}$. 
Further details will be explained in Sec.~\ref{sec5:analytical}.  
It is important that the $v$-$H$ relation has no higher-order corrections, 
because the DW at $c=0$ has only one length scale $\ell_0$.  

In the following sections, 
we will study DW dynamics driven by an external field $H$. 
We will calculate DW velocity in a stationary state 
and analyze how it depends on the coupling $c$ of multiple orders.  
We will also study how the internal structure in DW deforms in moving DW.


\section{Numerical analysis of DW velocity}
\label{sec3:Numerical}
\begin{figure}[tb]
\centering
\includegraphics[width=0.7\linewidth]{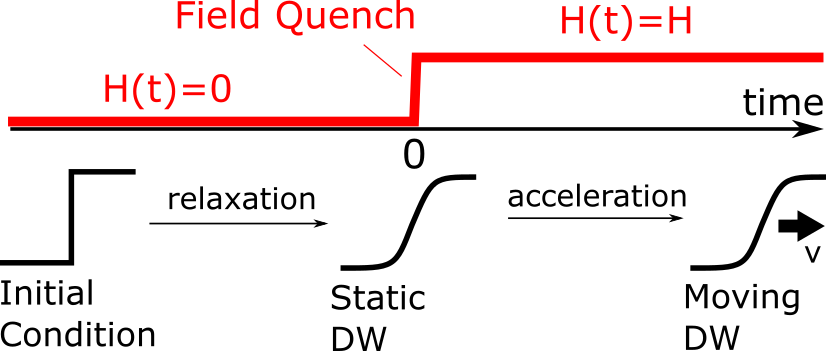}
\caption{
Simulation protocol and schedule of field control.}
\label{fig:fieldquench}
\end{figure}

In this section, we use numerical simulation
and study the dynamics of composite DWs driven by magnetic field $H$.
To this end, we numerically integrate the TDGL equations 
(\ref{eq:TDGL_F})-(\ref{eq:TDGL_P}) 
with the protocol shown in Fig.~\ref{fig:fieldquench} 
and calculate the velocity.  
Since the one-dimensional DW structure is concerned,
the Laplacian is replaced as $\nabla^2 \rightarrow \partial_x^2$. 
In the first part of the protocol, no field is applied and we integrate 
the equations Eqs.~(\ref{eq:TDGL_F})-(\ref{eq:TDGL_P}) 
to obtain a stable static DW solution.
This has a DW in two of the order parameters in zero magnetic field.  
In the second part, we apply a nonzero field $H$ and numerically 
track the time evolution of the order parameter fields 
until the DW velocity converges to a constant value.
Note that $v(c,H)$ has natural symmetry $v(c,-H) = -v(c,-H)$ 
as will be shown explicitly for the simplified model 
in Sec.~\ref{sec5:analytical}.  
Our choice for an initial state in the first part is
the configuration that left and right parts are 
different stable bulk states among the four
defined in Eqs.~(\ref{eq:stable_NF})-(\ref{eq:stable_P}).  
Chirality of DW, direction of spin rotation in xy-plane, 
is set in the initial condition.
We have calculated velocity of two types of composite DWs.  
One is ME-DW and $N_x$ and $P$ change their sign. 
The other is M-DW and $N_x$ and $n_x$ change sign. 
Note that P-DW does not couple to magnetic field. 

\begin{table}[tb]
 \caption{Model parameters in TDGL simulation}
\begin{tabular}{ll}
\hline
relaxation constants \ &
$\gamma =\gamma ' =\Gamma =1.0$
\\
stiffness &
$(\kappa, \, \kappa ', \, K )=(20, \, 4, \, 0.04)$
\\
anisotropy &
$\bm{\alpha}=(-0.5,-0.2,0)$
\\
Fe bulk term &
$(a, b)= (100,100)$
\\
Gd bulk term &
$(a', b')= (100,100)$
\\
P bulk term &
$A = 0.11$
\\
\hline
\end{tabular}
\label{table:simu_params}
\end{table} 

Let us summarize parameters used in the numerical simulations.  
The range $-100 \le x \le 100$ is used for the entire space
and discretized to $500$ grid points. 
The system size is thus $L=200$.  
Open boundary condition is used. 
As a DW moves, we follow its motion and 
shift the simulation region.  
The coupling constants in the free energy (\ref{eqn:free_ene_3D})
are set to the values listed in Table \ref{table:simu_params}.
Here, the identical value is set to all the relaxation 
constants for simplicity. 
Anisotropy parameter $\bm{\alpha}$ is set to describe 
experimental observation of magnetic moments. 
For stiffness constant, $K$ for polarization $P$ is set smaller 
than those for $\vNF$ and $\vNG$, 
since DW in ferroelectrics is generally much narrower 
than in magnets. 
In magnets it is typically the order of $\sim 100 \AA$ \cite{Kittel}, 
much larger than in BaTiO$_3$ \cite{Schilling2006}, 
a ferroelectric compound with the same structure with GdFeO${}_3$.

For numerical integration, 
we have used the fourth- and fifth-order 
Runge-Kutta-Fehrberg method \cite{Fehlberg1969}. 
We have used an adaptive discretization method for time 
so that the difference between the fourth and fifth order 
calculations is small enough for each order parameter. 
Typical time step is the order of $\Delta t \sim 0.01$

\begin{figure}[tb]
\centering
\includegraphics[width=0.8\hsize]{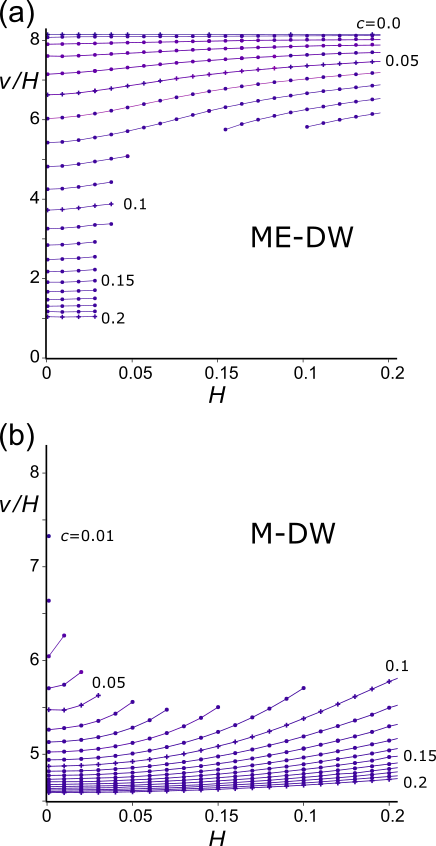}
\caption{
Domain wall velocity normalized by $H$ versus $H$.
(a) ME-DW and (b) M-DW.
The interaction $c$ changes in the range
$0 \le c \le 0.2$ by 0.01, but 
(b) has no data for $c=0$.  Minimum $H$ tested is $H=0,001$.
Note that the vertical axis has a different scale 
in (a) and (b).  
}
\label{fig:v_vs_H}
\end{figure}

When the coupling is $c=0$, the amplitude of Fe moment
is $N_0 = \sqrt{(100+0.5)/100}=1.0025$.
The mobility of its DW is calculated from Eq.~(\ref{eq:mu0})
\begin{equation}
  \mu_0 = 1.0025 \cdot \sqrt{20/0.3} = 8.185 . 
\end{equation}  

For $c > 0$, following the above procedures, we have calculated
the terminal velocity $v(c,H)$ in the stationary state
for ME-DW and M-DW with varying both magnetic field $H$
and the coupling $c$.
Figure \ref{fig:v_vs_H} shows the plots of $v(c,H)/H$ versus $H$
for $ 0 \le c \le 0.2$.  
The ratio $v/H$ is an effective DW mobility.  
Data are missing at some $H$-values when $c$ is large
for ME-DW and small for M-DW.
This is related to the instability of these composite DWs and we will
discuss this issue in the next section.  

The main characteristics of the effective mobility
$v(c,H)/H$ are as follows.  
First, this is a monotonic function
in both $H$ and $c$ but with opposite trends.  
The DW velocity and thus mobility slow down 
as the coupling $c$ increases.  
This agrees with an intuitive expectation that $c$ enhances 
dissipation, because it induces dragging the fields $\vNG$ and $P$,  
which do not directly couple to $H$.  
Concerning the $H$-dependence, 
the effective mobility grows monotonically with $H$.
It is important that it however never exceeds the value at $c=0$
\begin{equation}
  v(c,H) / H  \le \mu_0 , 
\end{equation}
for all the values of $c$ and $H$ examined.  
Secondly, ME-DW and M-DW have different characteristics,
particularly about $c$ dependence of velocity. 
We have continued calculation up to the largest value of $c=0.5$, 
and found that the effective mobility at $H \sim 0$ goes down 
to a small value $\sim 0.10$ for ME-DW. 
The scaling in the large-$c$ region is 
\begin{equation}
  v(c,H) / H \,  \bigr|_{H \sim 0} \sim 0.044 c^{-2}, 
\ \ \ ( 0.2 \le c \le 0.5 ) . 
 \end{equation}
In contrast to this, the mobility of M-DW remains a much larger 
value $\sim 4.5$ and shows a saturating behavior.  
It is noticeable that M-DW becomes unstable with $c \rightarrow 0$, 
but ME-DW is stable in that region and has a large mobility. 
The $H$-dependence in the small-$c$ region also differ in amplitude 
and more importantly in $c$-dependence. 

Let us discuss the origin of these differences between these two DWs.  
One important difference is in the continuity when switching on 
the interaction $c$.  
At small $c$, the ME-DW solution has small $|P(x)| \propto c^1$ near both ends 
of the system. 
This continuously evolves from the solution $P(x)=0$ at $c=0$. 
As for the M-DW solution, $n_x (x)$ is nearly $+n_0$ at one end of the system 
but $-n_0$ at the other end.  
This implies that the initial value at $c=0$ should be chosen as 
\begin{equation}
 \vNG (x) = \pm n_0 \,  
 \bigl( \sin \pi x / L , \pm \cos \pi x / L , 0 \bigr) . 
\end{equation}
Here, $L$ is the system size and very large (200 in our setting).
This large length scale appears because the angular part of Gd spins 
has a critical Gaussian form of the bulk free energy. 
As the system has no intrinsic length scale, the system size determines  
the spatial variation of $\vNG (x)$. 
When switching on the interaction $c$, the $\vNG$ field starts to 
couple to $\vNF$ and imports its length scale $\ell_0 \sim 8$. 
Therefore, the competition between the energy gain in the interaction 
part $F_1$ and the cost in the elastic energy of $\vNG$ field is serious, 
and the evolution of M-DW with $c$ is not as continuous as that of ME-DW.  
This yields qualitative different features in the $c$-dependence 
of $v(c,H)$ between the two types of composite DWs. 
We continue to analyze different features in more detail.  

\begin{figure}[tb]
\centering
\includegraphics[width=\linewidth]{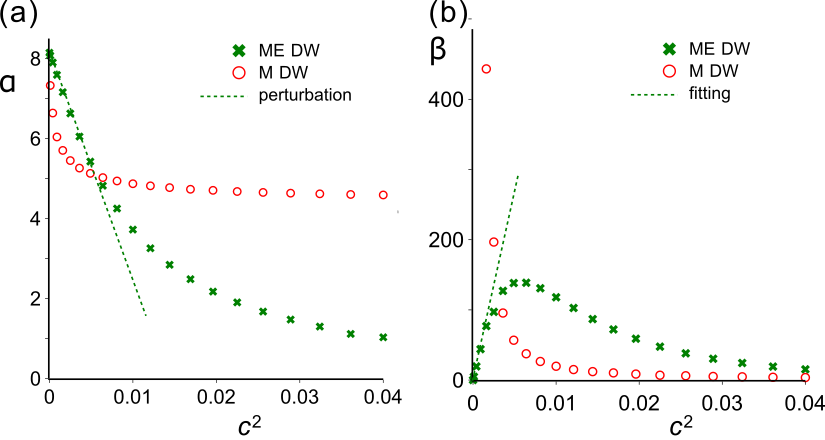}
\caption{
Linear and the third-order coefficients of DW velocity in driving field 
$v(c,H) = \alpha (c) H + \beta(c) H^3 + \cdots$. }
\label{fig:Speedcoeff_H}
\end{figure}

For analyzing $H$-dependence more quantitatively, let us expand the velocity 
in $H$ as 
\begin{equation}
  v(c,H) = \alpha (c) H + \beta (c) H^3 + \cdots . 
\label{eq:v_alpha}
\end{equation} 
The symmetry when $H$ is reversed ensures the relation 
$v (c, -H) = - v(c, H)$, as will be shown in 
Eq.~(\ref{eq:sym_H}) for a simplified model.
Therfore, even-order terms vanish in the expansion.  
By fitting the data in Fig.~\ref{fig:v_vs_H}, 
we have determined $\alpha(c)$ and $\beta(c)$ and plot the results 
versus $c^2$ in Fig.~\ref{fig:Speedcoeff_H}. 
$\alpha(c)$ and $\beta(c)$ are determined by fitting $v(c, H)$ 
in the range $H \le 0.10$. 
Note that data of M-DW for $c \le 0.04$ are missing, 
since the number of available data does not suffice 
due to DW instability.  
The linear coefficient $\alpha$ decreases monotonically with $c$ 
for both DWs, but it remains quite a large value $\sim 4.5$ for M-DW. 
These features are consistent with the trends 
in Fig.~\ref{fig:v_vs_H} explained before.  
In the small-$c$ region, the coefficient $\alpha (c)$ 
is well represented for ME-DW by 
\begin{equation}
 \alpha (c)  = \mu_0 + v_{21} c^2 + O(c^4 ) , 
\ \ \ v_{21} = -601.3 . 
\label{eq:alpha_v21}
\end{equation} 

The coefficient $\alpha (c)$ for M-DW also decreases with $c$ but 
$c$ dependence is more complicated. 
It shows a very quick decrease for small $c$ but then 
the decrease is strongly suppressed for $c > 0.01$.  

The difference is more prominent in the third-order coefficient $\beta (c)$. 
For ME-DW, it starts from 0 and shows a linear increase in $c^2$.  
The peak locates around $c=0.07$ and after that $\beta (c)$ decreases smoothly. 
The $c$-dependence is completely different for M-DW, and 
$\beta (c)$ shows a divergent behavior as $c$ approaches 0. 

In summary, it is a general characteristics that 
the DW mobility increases with $H$ and decreases with $c$, 
and this holds for both ME-DW and M-DW. 
However, analyticity of $c$-dependence differs between the two DWs. 
For ME-DW, the $c$-dependence is very smooth and can be represented 
by a power series. 
For M-DW, the $c$-dependence is continuous but cannot be 
represented by a simple power series in contrast to the case of ME-DW. 
This difference is attributed to a ``singular'' continuity 
of $\vNG$ field in M-DW upon switching on the interaction $c$. 


\section{Splitting instability of DW}
\label{sec4:instability}

\begin{figure*}[tb]
\centering
\includegraphics[width=0.85\linewidth]{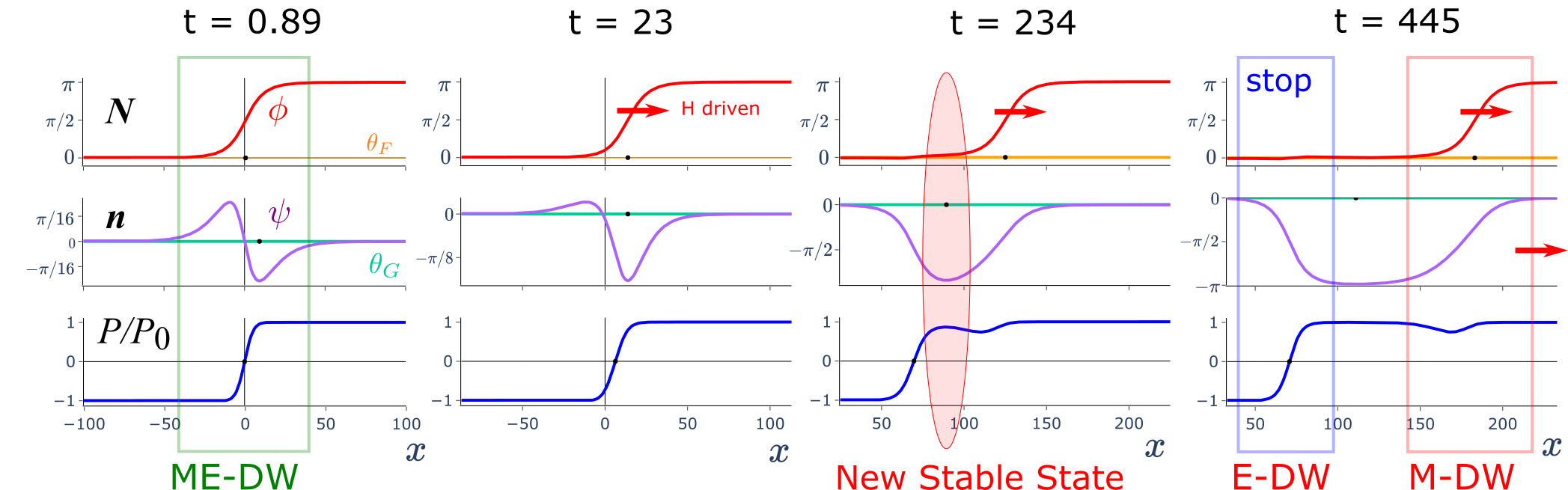}
\caption{Splitting of ME-DW into a pair of E-DW and M-DW. 
The panel for $t=0.89$ shows stable static profiles. 
$c=0.04$ and $H=0.10$.
}
\label{fig:splitprocess_ME}
\end{figure*}

\begin{figure*}[tb]
\centering
\includegraphics[width=0.85\linewidth]{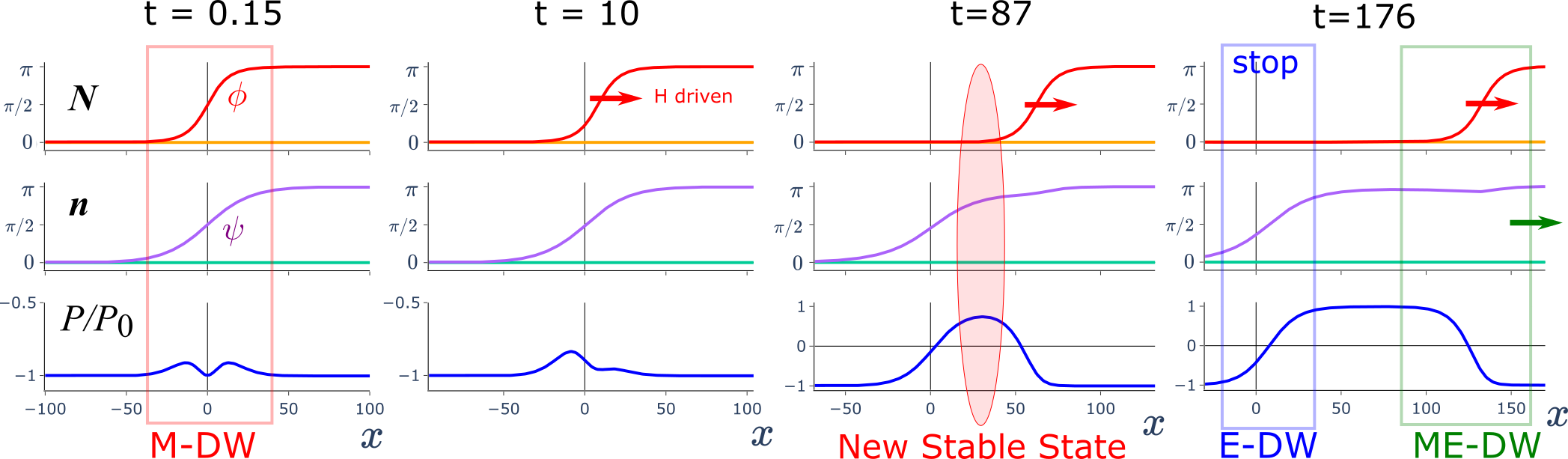}
\caption{Splitting of M-DW into a pair of E-DW and ME-DW. 
The panel for $t=0.15$ shows stable static profiles. 
$c=0.09$ and $H=0.10$.
}
\label{fig:splitprocess_M}
\end{figure*}

In this section, we discuss instability of composite DWs 
based on numerical results of their inner structure.   
We will show that the instability is a splitting into two composite DWs 
of different types. 
This splitting is unique in that the presence of multiple orders 
determines how it splits. 
The possibility of a similar splitting was pointed out 
in Tokunaga's work \cite{Tokunaga2009}
as impurity effects around a pinning center.
Our simulation has shown that splitting occurs in pure systems 
with no pinning centers. 
This means that the intrinsic inner structure of DW deforms 
and this leads a DW instability.  

To study the origin of DW instability, 
we follow the time evolution 
of the order parameter fields $\vNF (x,t)$, $\vNG (x,t)$, $P (x,t)$ 
and see what happens when a composite DW is unstable.   
Figures \ref{fig:splitprocess_ME} and \ref{fig:splitprocess_M} 
show a few snap shots of the field configurations for ME-DW and M-DW, 
respectively.
$(\phi, \theta_F)$ and $(\psi, \theta_G)$ are 
a set of polar and azimuthal angles of Fe and Gd spins, 
respectively in Eq.~(\ref{eqn:polarrep}). 
Polarization $P$ is plotted in the unit of $P_{0}=\abs{N_0n_0c/A}$.  
Length of spins $\abs{\bm{N}}$ and $\abs{\bm{n}}$ stays 
close to equilibrium value $N_0$ and $n_0$ in Eq.~(\ref{eqs:bulksols}), 
except a small drop typically $\sim 0.6$\% around DW center.  
Let us see the ME-DW case first.  
At a time short after the field quench ($t=0.89$), 
both $\vNF$ and $P$ fields have a nearly perfect DW and 
$\vNG$ field is slightly modulated around their center. 
As time goes ($t=23$ and 234), the two DWs are deformed 
and the DW in $P$ field delays.  
This delay is natural, since magnetic field drives $\vNF$ field 
and the drive to $P$ field is indirect through the coupling $c$.  
At the same time, the modulation in $\vNG$ is strongly enhanced 
around the DW in $P$ field and 
eventually comes close to $\vNG \sim (-1,0,0)$ around $x=90$ 
at $t=234$.  
In this $x$-region, $\vNF \sim (-1,0,0)$ and $P \sim 1$, and 
this set together with $\vNG \sim (-1,0,0)$ corresponds to 
one of the four stable bulk states (\ref{eq:stable_NF})-(\ref{eq:stable_P}). 
Once this stable region appears, its size expands with time 
as shown by the data at $t=445$.  
At the right end of this region, $\vNF$ and $\vNG$ flip 
constituting a M-DW and this moves to right. 
At the left end, $\vNG$ and $P$ flip constituting an E-DW and 
this stays at the position where it is created.  
Thus, one ME-DW is split into a pair of moving M-DW and unmoving E-DW. 
A similar process occurs for M-DW in Fig.~\ref{fig:splitprocess_M}. 
The DW in $\vNG$ field delays and $P$ modulation is enhanced around 
its center at the same time eventually to $P \sim 1$ around $x=30$. 
Then, this creates a E-DW and the right end of the expanding 
new stable region becomes a ME-DW.  
Thus, in this case one M-DW is split into a pair of moving ME-DW 
and unmoving E-DW. 

\begin{figure}[tb]
\centering
\includegraphics[width=0.9\linewidth]{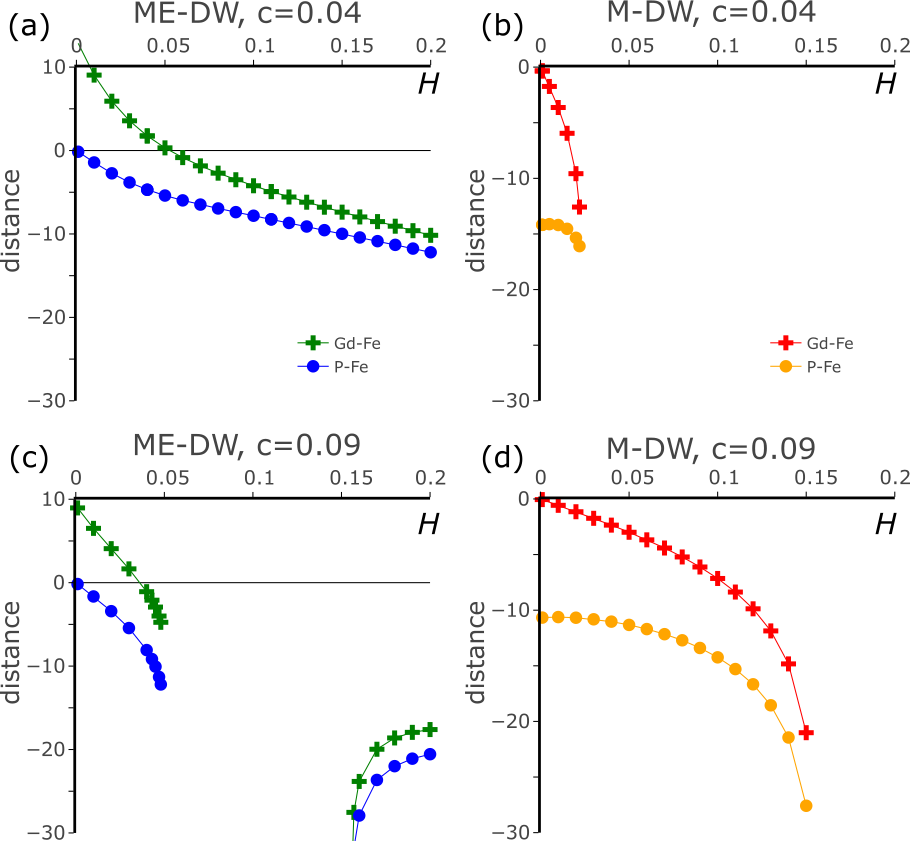}
\caption{Deformation of DW inner structure.}
\label{fig:innerstruct}
\end{figure}

To examine $c$-dependence more systematically, we have calculated 
the deformation of DW inner structure with varying $c$. 
Define a characteristic position $x_c$ for each order parameter field 
in a DW and plot their relative distances
in the stationary state 
in Fig.~\ref{fig:innerstruct} for $c=0.04$ and $0.09$.  
For each order parameter that flips across a domain wall, 
$x_c$ is defined by a point where its value is $\pi/2$ for 
$\phi$ and $\psi$ and 0 for $P$.
For a non-flipping order parameter, $x_c$ is defined 
by a point of maximal deviation from the value at $x=\pm\infty$. 

One can see two characteristic features in Fig.\ref{fig:innerstruct} 
with approaching the DW instability.  
First, the distances between $x_c (\mbox{Fe})$ and other two increase.  
However, the largest value just before instability depends on the value 
of $c$, and one cannot define a \emph{universal} critical 
value of the relative distances. 
Secondly, the two distances become closer with approaching the instability. 
This is due to two contributions.  One is that 
profile of non-flipping order parameters changes 
from a dispersive form to a single peak.  
The other is that the modulations in $\vNG$ and $P$ fields 
become more strongly coupled.  
It is interesting that the modulations in $\vNG$ and $P$ fields  
separate from $x_c (\mbox{Fe})$ by a distance 
a few times of $\ell_0$.  

\begin{figure}[tb]
\centering
\includegraphics[width=0.9\linewidth]{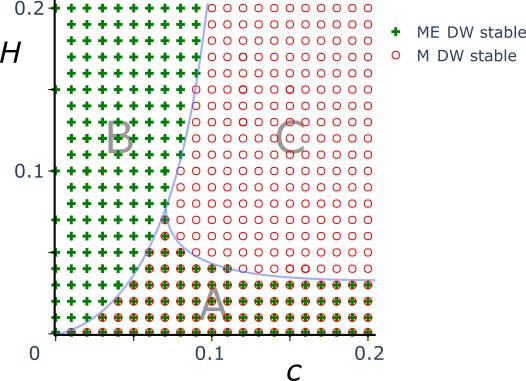}
\caption{
Phase diagram of DW stability. 
Green crosses and orange dots show the points 
where ME-DW and M-DW is stable, respectively. 
}
\label{fig:splitregion}
\end{figure}

Figure \ref{fig:splitregion} shows the non-equilibrium phase diagram of 
moving composite DWs in the $(c,H)$ parameter space.  
In region B, M-DW is unstable and splits into a pair of 
moving ME-DW and nonmoving E-DW, while 
ME-DW is stable. 
In region C, ME-DW is unstable and splits into a pair of 
moving M-DW and nonmoving E-DW, while M-DW is stable. 
Therefore, regions B and C are dual to each other.  
In region A, ME-DW or M-DW are both stable and 
this is only the region where one can drive both types of DWs 
without their splitting in applying magnetic field. 
One should note that its width is very small at small $c$  
but reasonably wide for $c \ge 0.7$.  
An interesting feature is a reentrant region $ 0.7 \le c \le 1.0$. 
In this region, increasing $H$ first destabilizes a ME-DW, and it
splits to a pair of M-DW and P-DW.  
However, increasing $H$ further beyond a second critical value  
now destabilizes a generated M-DW and a ME-DW comes back.  
Boundary of phase B is approximated 
as $H_c \propto c^k$ with $k = 1.8 \pm 0.1$ for the part with region A, 
and $k = 3.0 \pm 0.1$ for the part with C. 

This splitting instability is a result of deformation of DW inner 
structure, especially that of non-flipping order parameter. 
For ME-DW, increasing  $c$ and $H$ enhances deformation 
in $\psi$ around DW center. 
Once the peak value of $\psi$ comes close to $- \pi$, 
the order parameters around the peak position relax to another 
ground state with $\psi=-\pi$.  
This weakens clamping force between 
$x_c (\mbox{Fe})$ and $x_c (P)$ thus ME-DW splits. 
The same mechanism applies to M-DW and $P$ crossing 0 
is the condition in this case.

Two points are important in the splitting process. 
First, this splitting is a purely dynamical phenomenon, 
since generating a second DW costs an additional energy 
due to the gradient terms. 
Secondly, the inner structure of DW, especially the modulation of 
originally non-flipping order parameter, plays 
an important role in the process.                                                                                



\section{Analytical approach for dynamics of ME-DW}
\label{sec5:analytical}
In the last part, we use an analytical approach for studying 
the dynamics of ME-DW.  
The numerical results in the previous section show that 
the DW velocity $v(c,H)$ is a smooth function of both 
$c$ and $H$ near the decoupling limit $c=0$. 
We will perform a perturbative expansion for the solution 
of the TDGL equations and analytically determine the value of 
$v_{21}$ in Eqs.~(\ref{eq:v_alpha}) and (\ref{eq:alpha_v21}).

For simplicity, we make an approximation that 
the antiferromagnetic order parameters $\vNF$ and $\vNG$ are 
confined in the $xy$-plane and they do not change their amplitude
\begin{align}
 \vNF = N_0 (\cos \phiF , \sin \phiF , 0) , \ \ \ 
 \vNG = n_0 (\cos \phiG , \sin \phiG , 0) . 
\end{align}

This is justified if the anisotropy 
$\alpha_z - (\alpha_x + \alpha_y )/2$
 of Fe spins is large compared with $|c|$.
This simplification reduces the number of order parameter fields 
from seven ($\vNF$, $\vNG$, and $P$) to three ($\phiF$, $\phiG$, and $P$).  
As in Sec.~\ref{sec3:Numerical}, we study the case that the order parameters 
varies only along the $x$-direction.  
With the approximation above, the free energy density
(\ref{eqn:free_ene_3D}) is rewritten as 
\begin{align}
\mathcal{F}
&= \frac{\kappa}{2} ( \partial_x \phiF )^2 
- \frac{B}{4} 
\cos 2 \phiF - H \cos \phiF 
+ \frac{\kappa '}{2} ( \partial_x \phiG )^2 
\nonumber\\
&\phantom{=}
+ \frac{\kD}{2} ( \partial_x P )^2 + \frac{\aD}{2} P^2 
+ c \, P \, \cos ( \phiF - \phiG ) , 
\end{align}
where 
$ B \equiv N_0^2 ( \alpha_y - \alpha_x ) > 0$
and a constant is dropped.  
Here, the parameters are renormalized as: 
\begin{equation}
 N_0^2 \kappa \rightarrow \kappa , \ \ 
 n_0^2 \kappa' \rightarrow \kappa' , 
 N_0 H \rightarrow H , \ \ 
 N_0 n_0 c \rightarrow c . 
\end{equation}

Let us obtain a stationary state in which a DW travels with 
a constant velocity $v$ driven by an external field.  
To this end, it is convenient to rewrite the TDGL equation 
in the frame moving together with DW, and define the comoving 
coordinate $\xi = x -vt$. 
This comoving frame replaces time and space derivatives as 
$\partial_t \rightarrow -v \partial_\xi$ and
$\partial_x \rightarrow \partial_\xi$, 
and transforms the TDGL equations into a set of 
ordinary differential equations (ODE) in $\xi$.
It is important to note that the value of $v$ needs to be 
determined self-consistently.  
In this sense, the present task is similar to an eigenvalue 
problem of quantum mechanics in one dimension.  

The stationary TDGL equations thus calculated read as 
\begin{subequations} 
\begin{align}
- \frac{v}{\gamma} \dphiF  (\xi ) 
&=   
\kF \ddphiF  
+ \frac{B}{2}
\sin 2\phiF
- H \sin \phiF + c P \sin( \phiF - \phiG ),
\label{eq:ODE_F}
\\
- \frac{v}{\gamma '} \dphiG  (\xi ) 
&=   
\kG \ddphiG  - c P \sin ( \phiF - \phiG ) , 
\label{eq:ODE_G}
\\
- \frac{v}{\Gamma} \dot{P} (\xi )
&=  
\kP \ddot{P} - \aP P - c \cos ( \phiF - \phiG ) , 
\label{eq:ODE_P}
\end{align}
\label{eqn:XY-comoving-basic}
\end{subequations}
where a dot symbol denotes the derivative in $\xi$.  
Note that the system size is set infinite $-\infty < x < \infty$ 
in the following analysis. 

When $H$=0, uniform stationary states are easily calculated at, 
and there are four stable solutions corresponding to 
Eqs.~(\ref{eqs:bulksols}):
$(\phiF , \phiG , P)$ 
= 
$(0,0,-c/\aP )$, 
$(\pi,\pi,-c/\aP )$, 
$(0,\pi,+c/\aP )$, 
$(\pi,0,+c/\aP )$. 
Other solutions with $\phi = \pm \pi/2$ are unstable due to 
their large value of $\mathcal{F}$.  

For a ME-DW solution, we choose the following boundary conditions 
\begin{align}
&\phiF (-\infty ) = 0 , \ \ \phiF (+\infty ) = \pi ,  \ \ 
\phiG (\pm\infty ) = 0 ,  
\nonumber\\
&P (-\infty ) = -c/A , \ \ \  P (+\infty ) = +c/A .  
\label{eqs:ODE_BC}
\end{align}

For later use, let us examine how the sign change $c \rightarrow -c$ 
transforms a solution of Eqs.~(\ref{eqn:XY-comoving-basic}).  
It is easy to check that $(\phiF , \phiG , -P)$ is a solution for 
the coupling $-c$ with $v$ unchanged.  
This means that $P$ is an odd function of $c$ while $\phiF$, $\phiG$, 
and $v$ are even.  
The transformation upon field reverse $H \rightarrow -H$ is more 
nontrivial, but we have found the necessary transformation 
\begin{equation}
  \phiF (\xi ) \rightarrow \pi - \phiF (-\xi ) , \ \ 
  \phiG (\xi ) \rightarrow - \phiG (-\xi ) , \ \ 
   P (\xi ) \rightarrow  - P (-\xi ) . 
\label{eq:sym_H}
\end{equation}
and this also proves a natural expectation $v(c,-H)=-v(c,H)=-v(-c,H)$.  


\subsection{Case $c=0$}
\label{sec51:c0}
Let us summarize the result for a DW solution at $c=0$, which 
is a start of perturbation analysis. 
When $c=0$, the three order parameters are decoupled 
in Eq.~\ref{eqn:XY-comoving-basic}, and 
only $\phiF$ has a nontrivial solution, which is a DW.  
This is because spin anisotropy $B$ generates a characteristic 
length scale $\ell_0 = \sqrt{\kF / B}$ that determines a DW width.  
In case of no drive $H=0$, the stationary velocity is $v=0$, 
and Eq.~(\ref{eq:ODE_F}) is reduced to the time-independent sine-Gordon equation
$
 \ddphiF -  (2 \ell_0^2)^{-1} \sin 2 \phiF = 0 
$, 
and this has a static kink soliton solution 
$  
{\phiF}_s  (x ) = \pi / 2 \pm \sin^{-1} \tanh ( x / \ell_0 ) 
$. 
This describes a DW connecting two stable states 
$\phiF = 0$ and $\phiF = \pi$.  
Upon applying a driving field $H$, this DW moves 
and its motion is described by an exact solution 
found by Walker \cite{Schryer1974} 
\begin{equation}
 {\phiF}_s (\xi ) = \frac{\pi}{2} 
+ \sin^{-1} \left( 
 \tanh q \xi
\right) , \ \ \ 
q \equiv \ell_0^{-1} = \sqrt{\frac{B}{\kappa}} ,
\label{eq:Walker_sol}
\end{equation}
where the $+$ sign is chosen to satisfy 
the boundary conditions (\ref{eqs:ODE_BC}). 
Here, $\xi = x - v (H) t$ and the velocity is 
\begin{equation}
 v (H) = \mu_0 H , \ \ \ 
\mu_0 = 2 \gamma \left[ \int_{-\infty}^{\infty} 
\!\!\! d\xi \, \dphiF (\xi) ^2 \right]^{-1} 
= \frac{\gamma}{q} . 
\end{equation}
Thus, the velocity is linear in $H$ with no higher-order corrections. 
Its coefficient $\mu_0$ is a mobility of domain wall.


\subsection{Effects of $c \ne 0$}
\label{sec51:cne0}
Now, let us analyze the effects of the coupling $c$ on 
the velocity $v(c,H)$ of ME-DW driven by an external field 
using a perturbation approach. 
Following a standard procedure, 
we expand the velocity $v$ and the order parameters 
in Eqs.~(\ref{eq:ODE_F})-(\ref{eq:ODE_P})
in the coupling constant $c$ and driving field $H$. 
Recall the parity of the order parameters upon $c \rightarrow -c$, 
which was discussed before. 
This restricts the expansion in $c$ as 
\begin{subequations} 
\begin{align}
v 
&= \mu_0 H + 
\sum_{n=1}^{\infty} \sum_{m=0}^{\infty} v_{2n,2m+1} \, c^{2n} H^{2m+1} , 
\label{eq:exp_v}
\\
\phiF (\xi ) 
&= \phi_s (\xi ) + 
\sum_{n=1}^{\infty} \sum_{m=0}^{\infty} {\phiF}_{2n,m} (\xi) \, c^{2n} H^m , 
\label{eq:exp_F}
\\
\phiG (\xi ) 
&= 
\sum_{n=1}^{\infty} \sum_{m=0}^{\infty} {\phiG}_{2n,m} (\xi) \, c^{2n} H^m , 
\label{eq:exp_G}
\\
P (\xi ) 
&= 
\sum_{n=1}^{\infty} \sum_{m=0}^{\infty} P_{2n-1,m} (\xi) \, c^{2n-1} H^m . 
\label{eq:exp_P}
\end{align}
\label{eqs:exp_v-P}
\end{subequations}
The transformation (\ref{eq:sym_H}) upon $H \rightarrow -H$ implies that 
$\phiF_{2n m} (\xi )$, $\phiG_{2n m} (\xi )$, and $P_{2n+1 m} (\xi )$ 
are an odd function of $\xi$ for even $m$ and an even function for odd $m$.  

The expansion coefficients $v_{2n,2m+1}$, $\phiF_{2n,m}$, and others 
are to be determined successively from lower order to higher order. 

In the perturbative approach, we insert the expansions 
Eq.~(\ref{eqs:exp_v-P}) into Eqs.~(\ref{eqn:XY-comoving-basic}) and 
compare the terms of the same order $c^n H^m$ on both sides of the equations.  
Our goal is the first nontrivial correction to the velocity and 
that is $v_{21}$.  
We list up the lowest orders in the expansion up to $O(c^2 H)$, 
where $v_{21}$ first appears. 
\begin{subequations} 
\begin{alignat}{2}
&\mathrm{O} (c) : &
&\LP  P_{10} 
= 
- {\kP}^{-1} \, \tanh q \xi , 
\label{eq:pert10}
\\
&\mathrm{O} (cH): &
&\LP   P_{11} 
= 
- (\Gamma \kP )^{-1} \mu_0 \, \dot{P}_{10} , 
\label{eq:pert11}
\\
&\mathrm{O}(c^2): &
&\LF  \phiF _{20} 
= - {\kF}^{-1} P_{10} \, \sech q \xi , 
\label{eq:pert20}
\\ 
&\mathrm{O} (c^2H): \ &
&\LF  \phiF _{21} 
= - {\kF}^{-1} \bigl(\phiF _{20} \tanh q\xi + P_{11} \sech q\xi \bigr) 
\nonumber\\
&&
&\ \ \ 
-  (\gamma \kF )^{-1} 
\bigl( v_{21} q \sech q\xi + \mu_{0} \dphiF _{20} \bigr) , 
\label{eq:pert21}
\end{alignat}
\label{eqs:pert}
\end{subequations}
where we have used the relations 
$\cos \phiF _s (\xi ) = - \tanh q \xi$, and 
$\sin \phiF _s (\xi ) = \sech q \xi = q^{-1} \dphiF _{s} (\xi )$. 
Here, $\LP$ and $\LF$ are differential operators 
defined as 
\begin{subequations} 
\begin{align}
\LP  
&\equiv
\partial_x^2 - p^2 , 
\hspace{1cm} p \equiv \sqrt{A/K} , 
\\
\LF 
&\equiv
\partial_x^2 - q^2 \cos 2 \phi_s
= 
\partial_x^2 + q^2 ( 2 \sech ^2 q \xi - 1) 
\end{align}
\end{subequations}
We can solve the ODEs (\ref{eqs:pert}) by convolution with 
the Green function of the operators $\LP$ and $\LF$. 
$\LP$ is a modified Helmholtz operator, and its Green function 
$\GP$ is elementary.  
$\LF$ is a complicated operator but related to 
a quantum Hamiltonian for which all the eigenvalues and eigenfunctions 
are known \cite{GF1,GF2}.  
Using these results, we have calculated the corresponding 
Green function $\GF $.  
\begin{subequations} 
\begin{align}
2p \, \GP  (\xi ,\eta ) 
&= 
- e^{-p|\xi -\eta |} , 
\label{eq:Gq}
\\
4q \, \GF  (\xi ,\eta )
&=  
q |\xi -\eta | \sech q\xi  \sech q\eta  
\nonumber\\
&\ \ 
- e^{- q |\xi -\eta |}( 1 + \tanh q\xi  \tanh q\eta ) . 
\label{eq:Gp}
\end{align}
\end{subequations}
Note that these are symmetric: 
$\GP  (\xi , \eta ) = \GP  (\eta , \xi )$ and 
$\GF  (\xi , \eta ) = \GF  (\eta , \xi )$. 

We can solve equations (\ref{eq:pert10})-(\ref{eq:pert20}) 
successively using these Green functions: 
\begin{subequations} 
\begin{align}
P_{10} (\xi )
&= 
\kP ^{-1} 
 \GP  (\xi , \eta ) \circ \tanh q \eta , 
\label{eq:sol_P10}
\\
P_{11} (\xi )
&= 
(\Gamma \kP^{2})^{-1} \mu_{0} 
\GP (\xi , \eta ) \circ \partial_\eta 
[ \GP  (\eta , \zeta )\circ \tanh q \zeta ], 
\label{eq:sol_P11}
\\
\phiF _{20} (\xi )
&= 
(\kF \kP )^{-1} 
\GF (\xi , \eta )
 \circ [ \sech q \eta \, \GP  (\eta , \zeta ) \circ \tanh q \zeta ], 
\label{eq:sol_F20}
\end{align} 
\end{subequations}
where the circle symbol denotes a convolution 
$
  G_r (\xi , \eta ) \circ f (\eta ) \equiv 
  \int_{-\infty}^{\infty} \!\! d\eta \, G_r (\xi , \eta ) f(\eta), 
\ (r= p, \, q) 
$. 

We want to obtain $v_{21}$ and it appears in Eq.~(\ref{eq:pert21}). 
However, this equation contains a still unknown function $\phiF _{21}$. 
It is possible to obtain $v_{21}$ without calculating this.  
Notice that the operator $\LF $ has an eigenfunction with 
zero eigenvalue: 
\begin{equation}
 \LF \Psi_0 (\xi ) = 0, 
 \ \Rightarrow \  \Psi_0 (\xi ) = \sech q \xi . 
\end{equation}
Take an inner product between $\Psi_0$ and Eq.~(\ref{eq:pert21}), 
and then the left-hand side vanish because of the self-duality of $\LF $.  
Thus, 
\begin{align}
& 2 v_{21} 
= 
- \mu_0 q (\sech q \xi \, \tanh q \xi , \phiF _{20})
\nonumber\\
&\hspace{0.8cm} \ - \gamma
  \bigl( \sech q \xi , \phiF _{20} \tanh q \xi + P_{11} \sech q \xi \bigr) 
\nonumber\\
&= 
- \gamma  \bigl( \sech^2 q \xi , P_{11} \bigr) 
- 2 \gamma  \bigl( \sech q \xi \, \tanh q \xi , \phiF _{20} \bigr) .
\end{align}
where $(f,g) = \int_{-\infty}^{\infty}\!\! d\xi \, f(\xi ) g(\xi )$. 
We have used a partial integration for the term 
on the right-hand side in the first line. 
The value of $v_{21}$ is finally calculated as 
\begin{align}
&v_{21} = v_{21}^{(1)}+v_{21}^{(2)}, 
\\
&v_{21}^{(1)} 
= 
(8 p \Gamma \kP ^2 )^{-1} 
\kF \mu_0 
\int_{-\infty}^{\infty} \!\! d\xi \, \sech^2 q \xi 
\nonumber\\ 
& \times  
\int_{-\infty}^{\infty} \!\! d\eta \, 
e^{-p|\xi - \eta |} 
\int_{-\infty}^{\infty} \!\! d\zeta \, 
\sgn(\eta-\zeta)e^{-p|\eta - \zeta |} \tanh q \zeta , 
\\
&v_{21}^{(2)} 
= 
( p \Gamma \kP ^2 )^{-1} \kF \mu_0 
\int_{-\infty}^{\infty} \!\! d\xi \, \sech q \xi \tanh q \xi 
\nonumber\\ 
&\times  
\int_{-\infty}^{\infty} \!\! d\eta \, 
\GF  (\xi , \eta )  \sech q \eta  
\int_{-\infty}^{\infty} \!\! d\zeta \, 
e^{-p|\eta - \zeta |} \tanh q \zeta . 
\end{align}
We have performed this numerical integration and 
\begin{equation}
 v_{21} = -569
\end{equation} 
This agrees quite well with the fitting result 
in Sec.~\ref{sec3:Numerical}, 
$v_{21}^{\mathrm{fit}}=-601$. 

In this section, we have shown a perturbative calculation 
for the stationary dynamics of ME-DW driven by magnetic field. 
Our calculations justify the perturbative expansion of 
the terminal velocity $v(c,H)$ in both parameters $c$ and $H$, 
and provide a method for systematic improvements by including 
higher order terms. 
Expanded functions $\phi_{2n,m}$ and others show how the inner 
structure of ME-DW deforms in the stationary state from 
their static solution at $H=0$, and thus provide important 
information for understanding the DW dynamics.


\section{Conclusion}
\label{sec6:conc}
In this paper, we have studied the dynamics of composite
domain walls in a system with three coupled order parameters.
The system models the multiferroic material GdFeO$_3$, and
two antiferromagnetic order parameters of Fe and Gd spins
interact with Gd-ion displacement, which leads to electric polarization.
The main issue is how the interaction $c$ of the order parameters
changes the dynamics of DW.  
The multiferroic phase has three types of composite DW 
and we have studied two of them:
magneto-electric DW and magnetic DW. 
We have employed a phenomenological Ginzburg-Landau
model for describing this multiferroic system, 
and calculated how coupling $c$ is related 
to parameters of the microscopic spin Hamiltonian.

In Sec.\ref{sec3:Numerical}, 
we have numerically solved the corresponding time-dependent
Ginzburg-Landau equations when magnetic field $H$ is applied
to drive a DW and calculated the DW velocity $v$ in the stationary
state.
Analysis with varying $H$ and $c$ shows that
the effective DW mobility $v/H$  shows a monotonic
decrease with $c$ for both types of DW.
This is consistent with an intuitive picture that
the interaction $c$ provides a dragging of the order parameter 
fields not directly coupled to $H$, and that dissipates a DW motion. 
It also shows a monotonic increase with $H$ for both DW types
but the mobility never exceeds the value $\mu_0$ at $c=0$,
which corresponds to a non-composite DW in Fe antiferromagnetic order.  
However, its characteristics are quantitatively quite distinct
between the two types.
The mobility is a smooth function of both $H$ and $c$ for
magneto-electric DW, and we have confirmed its analyticity
by a perturbation theory in Sec.\ref{sec5:analytical} 
and calculated the leading correction in its $c$-dependence. 
This theory provides a formulation for systematic
calculation of further higher order terms of not only
DW velocity but also deformation in spatial structure
of the order parameters.  
In contrast to this, the mobility of magnetic DW
seems to depend on $c$ non-analytically. 

This is related to another interesting finding
discussed in Sec.\ref{sec4:instability}
That is a splitting instability of DW when driving magnetic
field exceeds a critical value.
A magneto-elastic DW splits into a pair of magnetic DW and
elastic DW, while a magnetic DW splits into a pair including
a magneto-elastic DW. 
One should note that the splitting occurs in bulk without impurity effects. 
An important difference between the two DW types is that
the splitting occurs in the large-$c$ region for
magneto-elastic DW but in the small-$c$ region for magnetic DW.
There is a narrow reentrant region where a magneto-elastic DW
becomes unstable for intermediate $H$ but
stabilized again at very large $H$.
Thus, magnetic DW is more singular than magneto-elastic DW
regarding its nonanalytic $c$-dependence of the mobility and
quick instability with $H$.
This is due to that fact that an evolution of magnetic DW upon switching
$c$ is not analytic in the part of Gd antiferromagnetic
order parameter.
It is characterized by a divergent length scale originating 
in the spin rotation symmetry at $c=0$, and
a finite length scale generated by the coupling to
Fe order parameter has a singular behavior in $c$.
To describe these nonanalytic characteristics qualitatively, we need
further analyses but leave them for a future study.  
  

\end{document}